\documentclass[4paper]{article}
\usepackage[utf8]{inputenc}
\usepackage[sort&compress,numbers,square]{natbib}
\usepackage[margin=1in]{geometry}
\usepackage{amsmath}
\usepackage{amsthm}
\usepackage{amssymb}
\usepackage{xcolor}
\usepackage{hyperref}
\hypersetup{
    colorlinks,
    linkcolor={red!50!black},
    citecolor={blue!50!black},
    urlcolor={blue!80!black}
}
\usepackage{graphicx, color}
\title{Bayesian Inference of Phenomenological EoS of Neutron Stars with Recent Observations}

\author{Emanuel V. Chimanski$^{1}$, Ronaldo
  V. Lobato$^{2,3}$, Andre R. Goncalves$^{4}$, Carlos
  A. Bertulani$^{2}$\\ \\
  $^{1}$National Nuclear Data Center, Brookhaven National Laboratory, Upton, NY, USA\\
$^{2}$Department of Physics and Astronomy, Texas A\&M University - Commerce, TX, USA\\
$^{3}$Universidad de los Andes, Bogot\'a, Colombia \\
$^{4}$Lawrence Livermore National Laboratory, Livermore, CA, USA
}

\begin{document}
\maketitle

\begin{abstract}
  The description of stellar interior remains as a big challenge for
  the nuclear astrophysics community. The consolidated knowledge is
  restricted to density regions around the saturation of hadronic matter $\rho _{0}
= 2.8\times 10^{14} {\rm\ g\ cm^{-3}}$, regimes where our nuclear
models are sucessfully applied. As one moves towards higher densities
and extreme conditions up to five to twenty times $\rho_{0}$, little
can be said about the microphysics of such objects. Here, we employ
a Markov Chain Monte Carlo (MCMC) strategy in order to acess the
variability of polytropic three-pircewised models for neutron star
equation of states. With a fixed description of the hadronic matter we
explore a variety of models for the high density regimes leading to
stellar masses up to $2.5\ M_{\odot}$. In addition, we also discuss
the use of a Bayesian power regression model with heteroscedastic
error. The set of EoS from the Laser Interferometer Gravitational-Wave
Observatory (LIGO) was used as inputs and treated as data
set for testing case.
\end{abstract}

\section{Introduction}

Neutron stars (NS) are supernova remnants, with a strong gravitational field and rapid rotation. They are objects with nuclear matter in one of the highest density states in the Universe. The matter in their interior is compacted to values from a few ${\rm g\ cm^{-3}}$ on their surface to possibly more than ${\rm 10^{15}\ g\ cm^{-3}}$ in their center.
The NS have become, alongside black holes, vital sources of gravitational waves, and although they
have been discovery more than 50 years as pulsars~\cite{hewish/1968}, its internal structure still is not thoroughly understood. Part of the challenge, relates to the extreme physical environments, e.g., large matter and energy densities, and the associated limits of our current models that contain parameters adjusted to reproduce, at their best, nuclear properties on natural conditions present on Earth.

 Recently, this picture has started to change with multimessenger observations~\cite{radice/2018a} from binary NS mergers~\cite{abbott/2017, abbott/2017a}. Those constraints provide the opportunity for a more detailed study about some of the parameters that describe global properties of NS such as radius constraints \cite{bauswein/2017, ligo/2018}, tidal deformabilities \cite{de/2018}, maximum mass \cite{margalit/2017} and other global properties. All this information, are intimately associated with the equation of state of the NS, and once one constrains the global properties, the microphysics can to be constrained as well. The GW170817 event, for example, besides the breakthrough of being the first gravitational wave detection, was also a source of many studies that considered the impact of the observation on internal aspects of the star. The impact of the NS crust on the equation of state was investigated \cite{gamba/2019}, as well as the effects of an isovector–scalar meson into the quark–meson coupling description of nuclear
matter \cite{motta/2019}, and also different Skyrme-like parametrizations \cite{lourenco/2020}. Non-parametric inference showed that the event favors soft EoS \cite{essick/2020}. Critical examinations of the EoS of dense matter were performed \cite{tews/2018} considering the nuclear physics in the chiral effective field theory framework, but still left some understanding to be improved in regions of high densities of the EoS. The association with electromagnetic counterparts of the event, lead to the first time to a joint-constraint. Using the binary's tidal deformability parameter, simulations of EM observations within numerical relativity and Kilonova models, extreme EoS models were ruled out, theoretically the stiffest and softest ones, e.g., see figure 2 of Ref. \cite{radice/2018a}. Statistical Bayesian methods were applied in the context of the GW170817 event, where microscopic models of cold neutron stars using chiral effective models \cite{lim/2018} were studied. Recently, the GW event with X-ray sources we combined and studied with the relativistic mean field models \cite{traversi/2020}. Besides the electromagnetic counterpart of the binary merger, another important recent electromagnetic measurement was done by the NASA's {\it Neutron Star Interior Composition Explorer} ({\it NICER}) \cite{gendreau/2016}, also constraining the mass-radius of the pulsar PSR J0030+0451 \cite{miller/2019a, riley/2019}. While the astronomical data was gathered and studied theoretically, experiments on Earth have also been performed. For example, the Lead Radius EXperiment (PREX-2) which has provided a better understanding of the nuclear matter around the saturation density, has a direct implication for the crust of neutron stars. The extrapolation of the data to higher densities has limited the stellar radii to $13.25 \lessapprox R_{1.4} \lessapprox 14.26$ km, meaning that the EoS should have a softening in the intermediate region and a stiffening at the high densities. This, in turn, could lead to a phase transition in the stellar core. The increment in observational data, has helped to establish further constraints on the dense matter EoS opening a rich field for statistical and machine learning models \cite{fujimoto/2020, morawski/2020, krastev/2021, soma/2022a, krastev/2022}.

The description of nuclear matter around the nuclear saturation density $\rho _{0} = 2.8\times 10^{14} {\rm\ g\ cm^{-3} = 0.17\ fm^{-3}}$ is well understood in terms of hadron physics. The microphysics at intermediate densities is yet far away from a consensus with a wide range of possible models. The debate includes the binding nature of NS with theories considering self-bound quarks or simply with gravity-bound systems. The asymptotic behavior of the EoS, on the other hand, has been understood in the context of quark matter \cite{Alford-2008}. As the details of the nuclear models are out of the scope of this work, we refer to Refs~\cite{Weber-2005,Kurkela-2014,Gorda-2018,Zhou-2018,Li-2020,Miller-2019,Sedaghat-2021,Miao-2021,Traversi-2022} and references within for more information.

In this work, we separate the description of the equation of state into a three-piece polytropic functional. We based our approach in the work by Read et al. 2009 \cite{read/2009} where a piecewise EoS was fitted with a direct cost function minimization. Here, we extend this picture to a larger class of models made possible with modern computing resources. We adjust the position of each piece of the EoS to better reproduce the observational data, and then perform a Bayesian Inference with Markov Chain Monte Carlo on the polytropic exponent of each case. This approach provides an assessment of the impact of variations in the EoS at intermediate and high densities on the mass radius diagram of the star.

One of our objectives is to determine the mass and radius of a
selection of stars in correlation to the description of nuclear matter
modeled by the EoS. In this way, we can systematically use different
EoS parametrizations to determine relevant characteristics of neutron
stars. In addition to that, we discuss the use of a Bayesian
statistical model with heteroscedastic errors. This enables the
training of statistical models based on EoS generated by different
nuclear physics pictures. Due to the various parametrizations present
in the microscopic models, the result set of all equation of states
has a variance that increases alongside density
(heteroscedasticity). This behavior can be captured by models with
scattering residuals at different levels of the EoS when trained
simultaneously with the NumPyro probabilistic programming
library\footnote{\url{https://num.pyro.ai}}. Here we use the set of
EoS from the Laser Interferometer Gravitational-Wave
Observatory (LIGO) as input and handling the data set as a test case.

\section{The structure of Neutron Stars}
The description of NS comprises both the quantum mechanical and general relativity worlds. The properties of particles that constitute the stellar matter are considered via equation of state obtained from quantum mechanics in flat space. The EoS is present in the energy-momentum tensor
$T^{\mu \nu}(\rho, P(\rho))$, the bridge to the gravitational/geometric degrees of freedom $G^{\mu\nu}$, through Einstein’s general relativity equations
\begin{equation}
  G^{\mu\nu} \equiv R^{\mu\nu} - \frac{1}{2}g^{\mu\nu}R = 8\pi
  T^{\mu\nu}.
\end{equation}

For a perfect fluid energy-momentum tensor and for a static spherical symmetric spacetime, the Einstein's field equations lead to the hydrostatic equilibrium equation, well-known as
Tolman-Oppenheimer-Volkoff equation~\cite{tolman/1939,
  oppenheimer/1939}. This equation reads in natural units
\begin{equation}
\label{TOV-Eq}
  p' = - (\rho + p) \frac{4\pi pr + m/r^2}{(1 - 2m/r)},
\end{equation}
where the prime indicates radial derivative and $m$ is the gravitational mass enclosed within the
surface of radius, i.e.,
\begin{equation}
  m' = 4\pi\rho r^2.
\end{equation}
To solve this system one needs to add to it an EoS ($p(\rho)$) and use the boundary conditions
\begin{equation}
m(r)|_{r=0}= 0,\quad p(r)|_{r=0}=p_{c}\quad \textrm{and}\quad \rho (r) |_{r=0}=\rho _{c},
\end{equation}
where $p_c$ and $\rho_c$ are the pressure and density at the center of the star. The numerical integration of Eq.~(\ref{TOV-Eq}) follows the pressure decrease as one moves away from the center, and it is stopped when the condition
\begin{equation}
p(r)|_{r=R}=0
\end{equation}
is reached at the surface of the star $R$. The integration of the profile density
\begin{equation}
M(R) \equiv 4 \pi \int_0^R r^2 \rho(r) dr
\end{equation}
provides the total gravitational mass of the star $M$.  The resulting
M-R relation can be compared to data from astronomical observations. Once the EoS is provided, the global properties of the
neutron stars can be obtained. However, until recently the uncertainties
in the mass-radius relationship were significantly large so that almost any EoS could describe the same stellar structure.

The NS can be subdivided in many layers with different
theories. Roughly, we can have four regions for the interior: the inner
and outer core and the inner and outer crust. For the
exterior part, an atmosphere with plasma governed by strong magnetic/electric fields is frequently assumed. The theories to describe the interior span many-body theories of high dense strongly interacting systems, nuclear many-body theories in the high
density-temperature regime, atomic structure and plasma physics,
respectively \cite{haensel/2007}. We recall that due to all these different regimes/densities, only the outer crust is well understood, since one can compare with experimental data of atomic nuclei. Around
the nuclear saturation density and above, the constraints become too fragile allowing for many descriptions of the NS interior: for the outer
core $npe\mu$ (neutron-proton-electron-muon) plasma and for the inner
core many possibilities such as fermion/boson condensation, hyperons, pion/kaon condensation, strange quarks surrounded by hadronic matter and so on. This complex puzzle, calls for an extension of our
knowledge about the many-body physics regimes and should lead to
models able to describe a large variety of environments all at once.

\section{The equation of state}
The description of the outer crust inside neutron stars is well accepted to be given in terms of hydronic matter up to the saturation density $\rho _{0} = 2.8\times 10^{14} {\rm\ g\ cm^{-3}}$. This limit reflects the
validity of well established nuclear structure models that were developed to describe properties of heavy atomic nuclei on Earth. When one goes beyond $\rho _{0}$, more sophisticated degrees of freedom, as mentioned in the previous section, have to be considered. These extra variables make a universal and simultaneously description of systems with such large range density profiles a challenging task.
The microscopic constraints are so far just a few and consist of $dp/d\rho$ being always positive and well-defined with $p\ge 0$, electric neutrality, beta equilibrium and the speed of the sound must be less than the speed of light.

Generally speaking, the different set of EoS can be separated accordingly to the compressibility of the nuclear matter: soft and stiff and the corresponding speed of sound. Among the several microscopic methods for the EoS we cite: Perturbation expansions within the
Brueckner-Bethe-Goldstone theory, perturbation expansions within the Green's-function theory,
variational methods, effective energy-density functionals, and relativistic mean-field (RMF)
models~\cite{bethe/1971, ring/1980, blaizot/1985, machleidt/1989, akmal/1997,
  haensel/2007}. Point-coupling and non-relativistic models employing well known nuclear interaction such as Skyrme and Gogny are also
used~\cite{nikolaus/1992, friar/1996, skyrme/1958,
  bell/1956, skyrme/2006, decharge/1980, berger/1991}. Two approaches
are frequently seen in the literature: Models that approach the
physics around $\rho _{0}$~\cite{dutra/2012, dutra/2014,
  lourenco/2019}; or models that aim specific systems such as binary
neutron star mergers, e.g. using LIGO-VIRGO observational data for the
mass-radius of NS to extract the embedded EoS model
\cite{Miao_2021}. In general, the EoS are generated through these
models using parameters adjusted to reproduce fundamental physical quantities and are listed in tabulated data, i.e., there are many models and
many codes/ways to generate them. The
phenomenological models have the advantage of being easily parametrized and can generate EoS that reproduce the M-R diagrams, offering simpler representations of sophistical microscopic calculations. These are the so-called
representation of the EoS, which are basically two: the
piecewise-polytropic \cite{mueller/1985, read/2009, hebeler/2010, steiner/2010,
  ozel/2016a, raithel/2016} and spectral representations
\cite{lindblom/2012, lindblom/2014}. Here we focus on models of the first kind.

\subsection*{Piecewise-polytropic representation}
The piecewise polytropic model consists of a connected set of polytropic equations, effectively power-law like functions, with different exponents (also called indices) to account for the softness/stiffness of the EoS at a given density regime. The indices are free parameters in most
of the cases when one considers this kind of parametrization. The density where the transition between two polytropic equations take place can also be used as a free parameter specially at highly dense regions
\cite{ozel/2009}. The polytropic representation can yield macroscopic observables for a wide range of
EoS with only a few parameters. The stellar structure maps the EoS parameters to gravitational
mass, radius, moment of inertia and others global properties. This
representation has been extensively used in NS studies, gravitational waves
simulations \cite{lackey/2015, carney/2018, ma/2018, east/2019} and
can be tested using the astronomical data such X-ray, gamma and
gravitational waveforms. The representations can be also very useful
when dealing with modified gravity such as $f(R)$
\cite{astashenok/2013, teppapannia/2017} and or other alternative theories where a coupling between
geometry and matter could introduce corrections in the energy density, and
therefore requires an analytical representation to model the stellar structure \cite{lobato/2020}.

\begin{figure}[t!]
\centering
\includegraphics[width=10. cm]{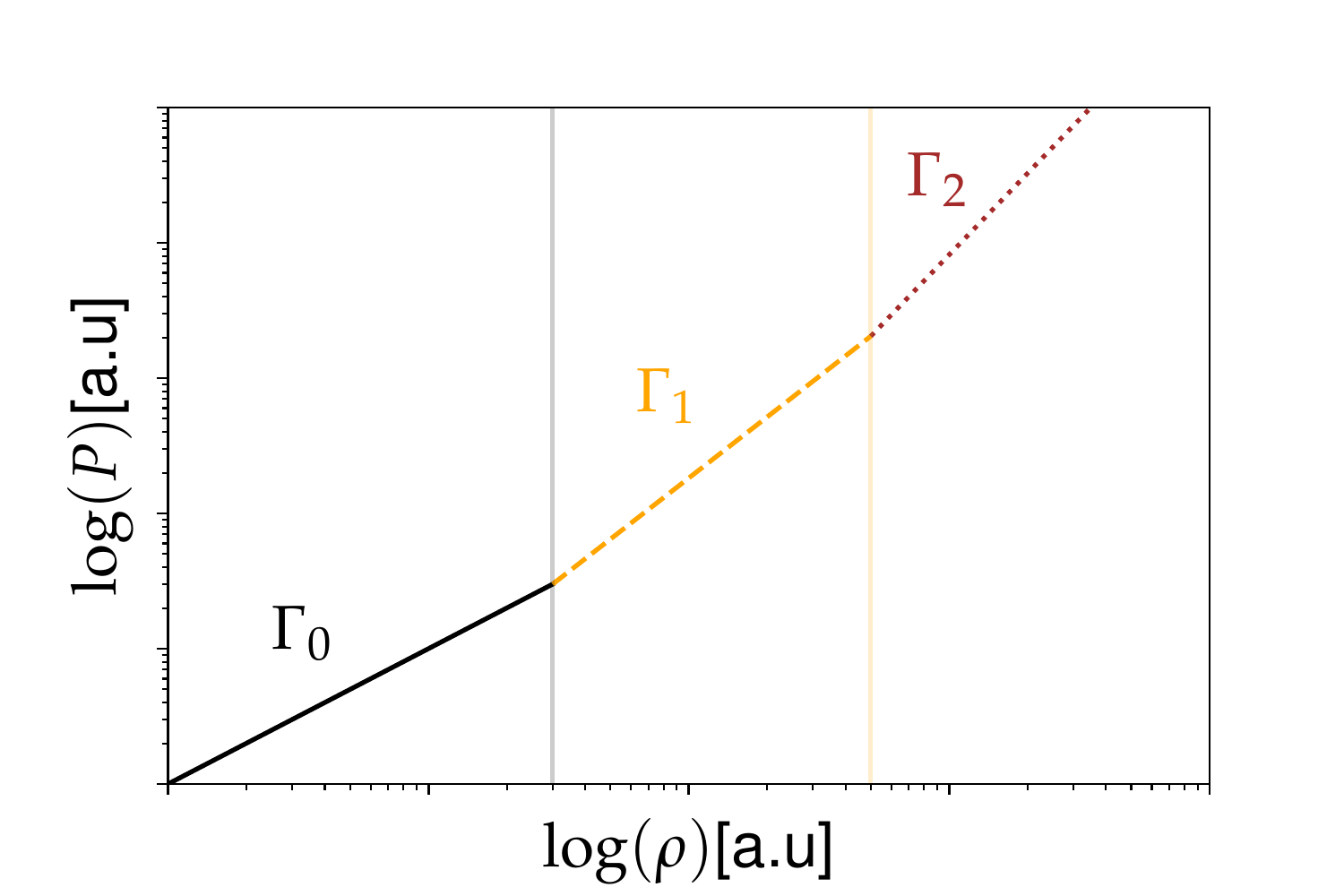}
\caption{Piecewise model representation of the equations of state with the
  polytropic equations (\ref{Poly}). The vertical lines represent the
  transition points $\rho _{0\rightarrow 1}$ and $\rho _{1\rightarrow 2}$ of each piece of the EoS.\label{piecewise_model}}
\end{figure}

The piecewise-polytropic parametrization of the EoS can be written as \cite{read/2009}
\begin{equation}
\label{Poly}
    P(\rho) = K_i\rho ^{\Gamma_i},
\end{equation}
where $\Gamma_i$ are the polytropic indices and $K_i$ strength constants. Due to the continuity of pressure at the transition points, we impose (with $i>0$)
\begin{equation}
K_{i} = K_{i-1}
\rho_{t}^{\Gamma_{i-1}-\Gamma_{i}}.
\label{ks}
\end{equation}
In this work the piecewise polytropic parametrization of the EoS is done
by combining three different polytropes. We have two parameters to set the transitions points and three values of $\Gamma$ for each region. For
the first piecewise, i.e, at density values smaller than
$\rho_{0\rightarrow 1}$, see the gray vertical line in
Fig. \ref{piecewise_model}, we define $\Gamma_0$
(and $K_0$) in connection to the SLy4 EoS in this regime. This equation of state, describes very well the nuclear matter and match the BPS and HP94
based on experimental nuclear data, e.g., see Fig. 1 of Ref. \cite{haensel/2004}. We set $\Gamma_0 = 1.475$
and $K_0 = 1.475\times 10^{-3}\ {\rm[fm^3/MeV]^{2/3}}$.

The two other polytropic combination of (\ref{Poly}) have the set
of parameters $\{\Gamma _{1}, \Gamma _{2}\}$, and the respective transition
taking place at $\rho _{0\rightarrow 1}$ (see
Fig.~\ref{piecewise_model}). Having the first part of the EoS fixed the transition point $\rho_{1\rightarrow 2}$ is placed at two different density values and the exponents $\{\Gamma_1, \Gamma_2\}$ are analyzed with statistical methods in relation to data from astronomical observations.

\section{Markov Chain Monte Carlo and Bayesian Inference}
Markov Chain Monte Carlo (MCMC) is a convenient numerical way to stochastically explore a space of parameter values with high probability and provides good expectation estimates for model variability. This is basically the point of any Bayesian inference quantification summarized in terms of mean and variance values. One assumes a distribution $F$ for a given parameter with mean value $q_{t}$, an associated uncertainty $\sigma _{t}$, and with a transition probability $K(q'|q)$ we can write
\begin{equation}
F(q) = \int dq' F(q') K(q'|q).
\end{equation}

 Since the form of $F$ is preserved, the algorithm can start from any point $q'$ that the convergence to the typical parameter space region is guaranteed. Sampling from a prior distribution $F$ and employing a simple Metropolis algorithm with $t=0,\ldots, t=T$ iterations to construct the Markov Chain $[q_{t=0},q_{t=1},\ldots ,q_{t=T}]$ one approximates the posterior distribution given a sufficient large number of steps $T$. The sampled value $q$ is accepted according to the probability
 \begin{equation}
 \Pi = \frac{f(\chi ^{2} (q_{t+1}))}{f(\chi ^{2} (q_{t}))},
 \end{equation}
where $f(x)=e^{-x}$ and $\chi ^{2}$ are the likelihood and chi-squared functions, respectively.
 More details and algorithms can be found in Refs~\cite{Kroese-2011,Gill-2014,Betancourt-2017}.

 Here, we employ the MCMC algorithm to access the variability of polytropic
 piecewise-like models. The set of parameters is assumed to be
 uncorrelated and normally distributed. We defined 5 (MD$\#$) models containing
 different parametrizations schemes. The parameters of the first
 polytrope (left side of the gray vertical line of Fig.~\ref{piecewise_model}) is kept unchanged
 with $\rho _{0} \equiv \rho _{0\rightarrow 1}= 2.8\times 10^{14} {\rm\ g\
   cm^{-3}} = 0.17$ fm$^{-3}$ and $\Gamma _{0} = 1.475$. This low
 density region of the EoS is well understood in terms of hadronic
 matter. For densities values $\rho _{0\rightarrow 1} < \rho \le
 \rho_{1\rightarrow 2}$ we use the second polytrope, representing an
 intermediate high-density portion of the EoS with adiabatic index $\Gamma
 _{1}$. This segment of the EoS (center part of
 Fig.~\ref{piecewise_model}) is not fully understood with our current
 knowledge of microphysics, representing a density region where EoS
 variability can be studied. Finally, the transition to the third
 polytrope with adiabatic index $\Gamma_2$ (right side Fig.~\ref{piecewise_model})) is defined by the
 density transition region $\rho _{1\rightarrow 2}$, representing the
 densest part of the EoS. The value defining the transitions is difficult to be estimated since the physics of highly dense interacting matter is yet not known in details, and therefore it can be arbitrarily chosen.

 We worked with two different values for the transition region $\rho _{1\rightarrow 2} = \{4.8\rho_0;
 7.2\rho_0\}$. These two choices will have the two following
 sets of polytropic indices $\Gamma_1 = \{2.5, 2.6, 2.8, 3.0\}$ and
 $\Gamma_2 = \{1.8, 1.9, 3.0, 3.3, 3.7\}$ for the polytrope 1 and 2, respectively. Once the transition
 point $\rho _{1\rightarrow 2} $, and $\Gamma_{1,2}$ are defined, one can calculate the
 constant $K$ for each individual polytrope using Eq. \eqref{ks} and have the full the description of the EoS.

 In total, we have 5 combinations of parameters
 schemes. Our MD$\#$s, provide a large variety of equation of states,
 i.e. different densities and pressure profiles to be used in solving the TOV
 equation, and generating the respective stellar global properties. Our models are
 summarized in Table~\ref{table_MCMC}, where the
 color scheme used in the figures is also provided.

\begin{table}[ht]
\centering
\begin{tabular}[t]{lllll}
\hline
Label	& $\rho _{1\rightarrow 2}$ & $\Gamma _{1}$ & $\Gamma _{2}$ & Color \\
\hline
\hline
MD1 & $7.2\rho_0$ & 2.5 & 1.8 & cyan\\
MD2 & $7.2\rho_0$ & 2.6 & 1.9 & pink\\
MD3 & $4.8\rho_0$ & 2.6 & 3.0 & purple \\
MD4 & $4.8\rho_0$ & 2.8 & 3.3 & brown\\
MD5 & $4.8\rho_0$ & 3.0 & 3.7 & lime\\
\hline
\end{tabular}
\caption{\label{table_MCMC}Summary of the parameters for piecewise EoS
  models. $\rho _{0} = 2.8\times 10^{14} {\rm\ g\ cm^{-3} = 0.17\ fm^{-3}}$ is the nuclear saturation density. We assumed
  the uncertainty as $\sigma = 0.01$ for both $\Gamma _{1,2}$ in each model.}
\end{table}

 The MCMC is applied for $\Gamma _{1}$ and $\Gamma _{2}$ with an error
 uncertainty of $\sigma = 0.01$ associated to each of the exponents. All the other parameters are kept
 fixed. Small variations in the polytropic indices lead to large change in the pressure, specially at density regimes of $>4\rho _{0}$. Therefore, the variation of $\Gamma _{1,2}$ provides a way to access the impact of EoS variability on the mass radius diagram for the models employed here.

\begin{figure}[h!]
\begin{center}
\includegraphics[scale=0.4]{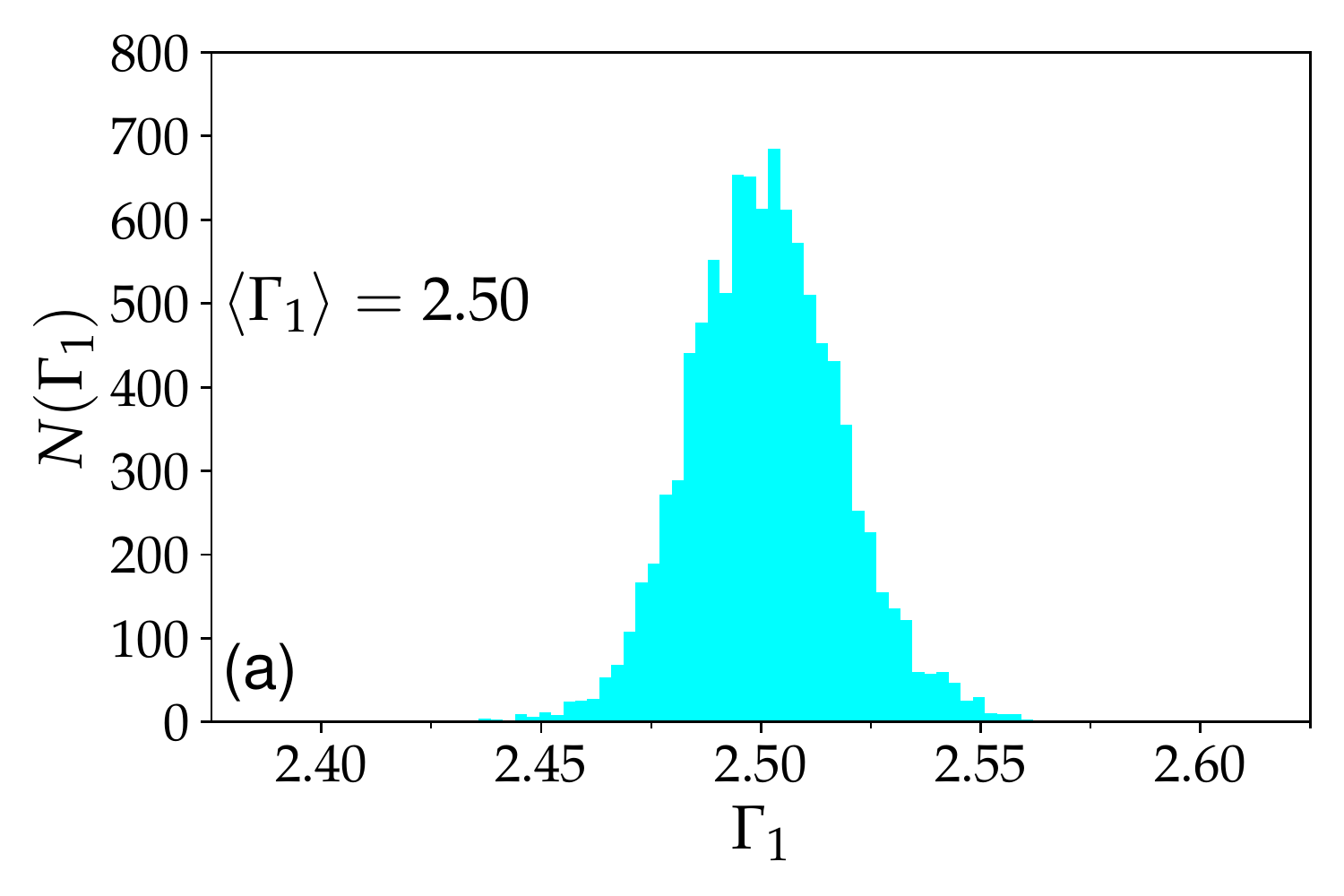}
\includegraphics[scale=0.4]{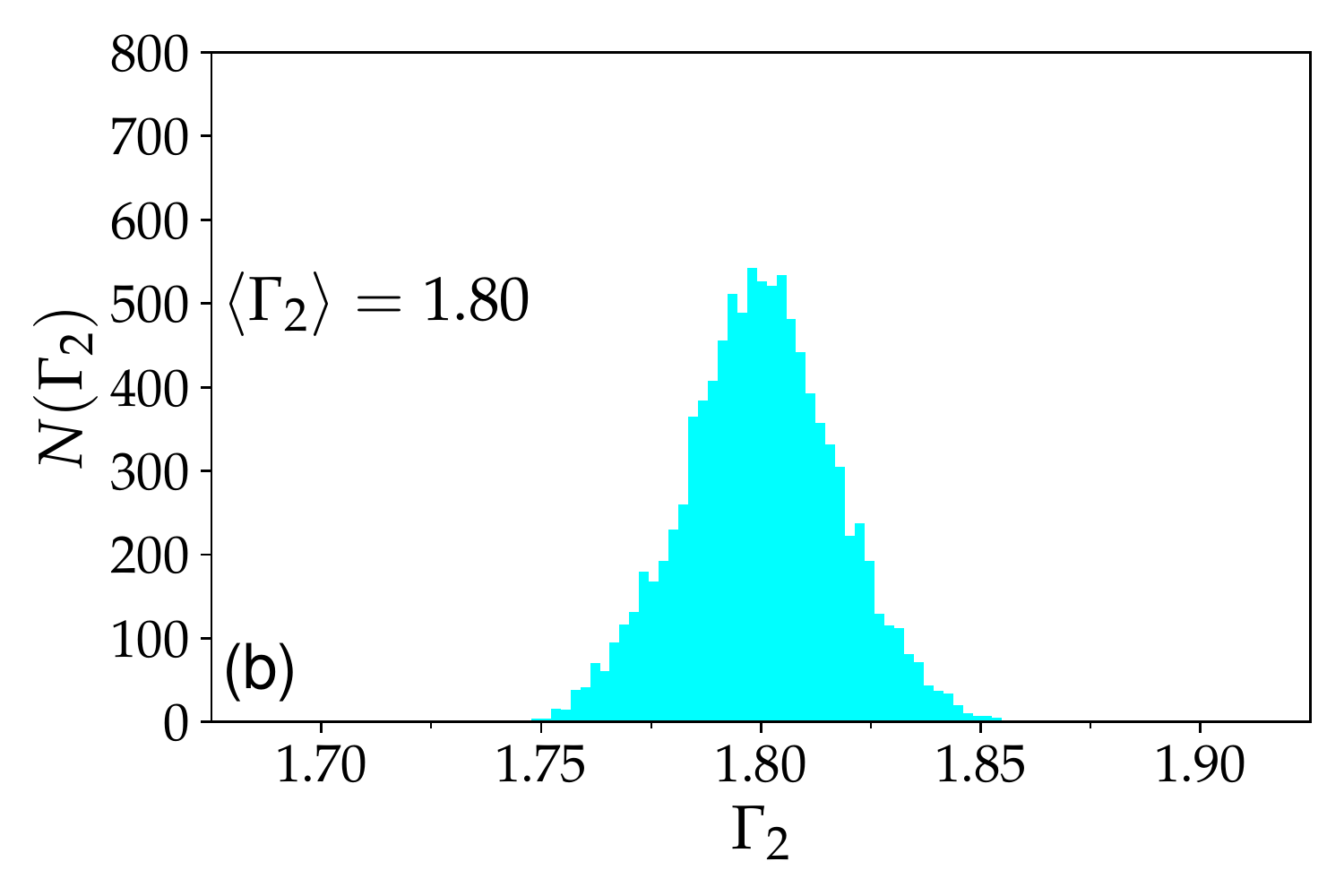}\\
\vspace{-0.7cm}
\includegraphics[scale=0.4]{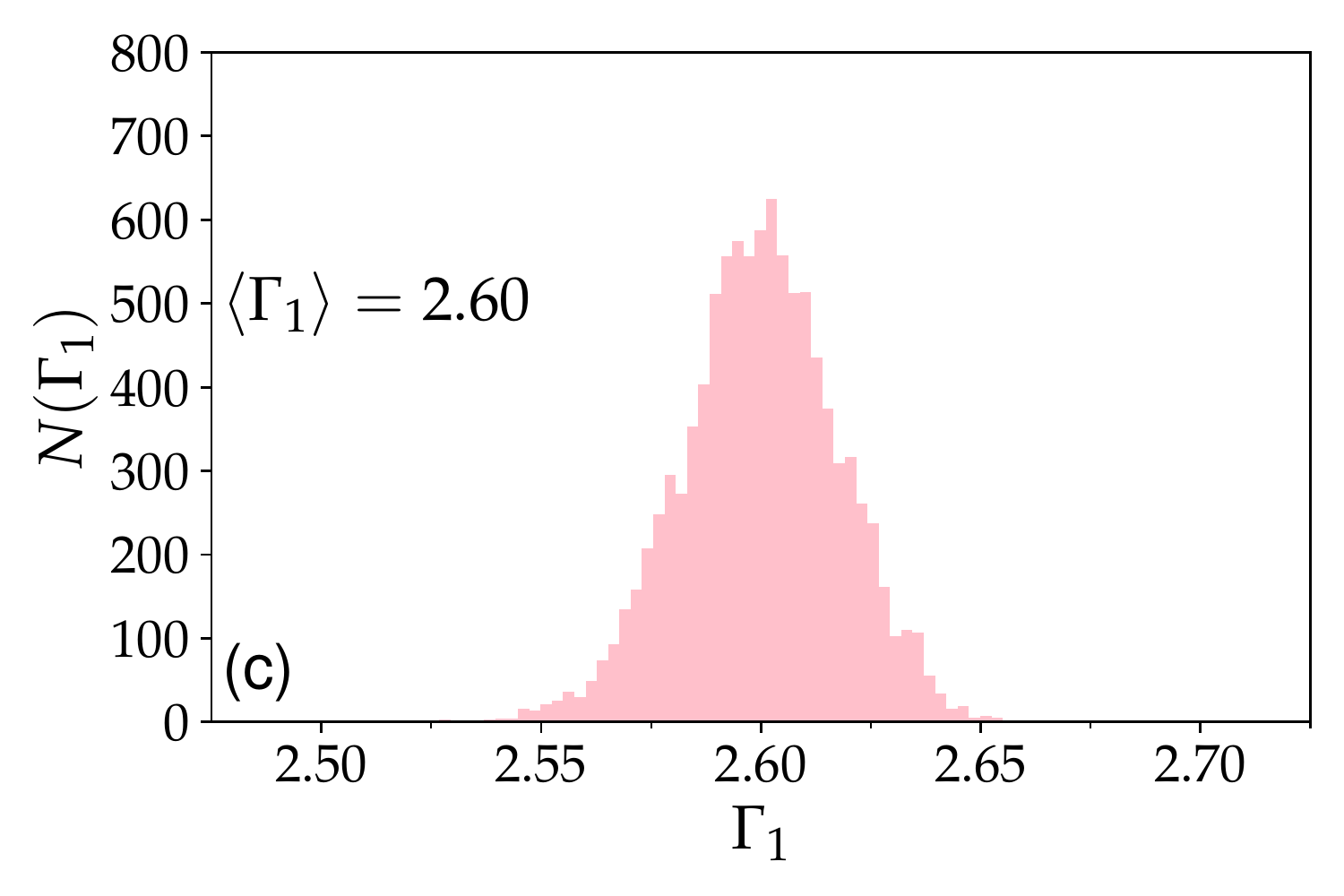}
\includegraphics[scale=0.4]{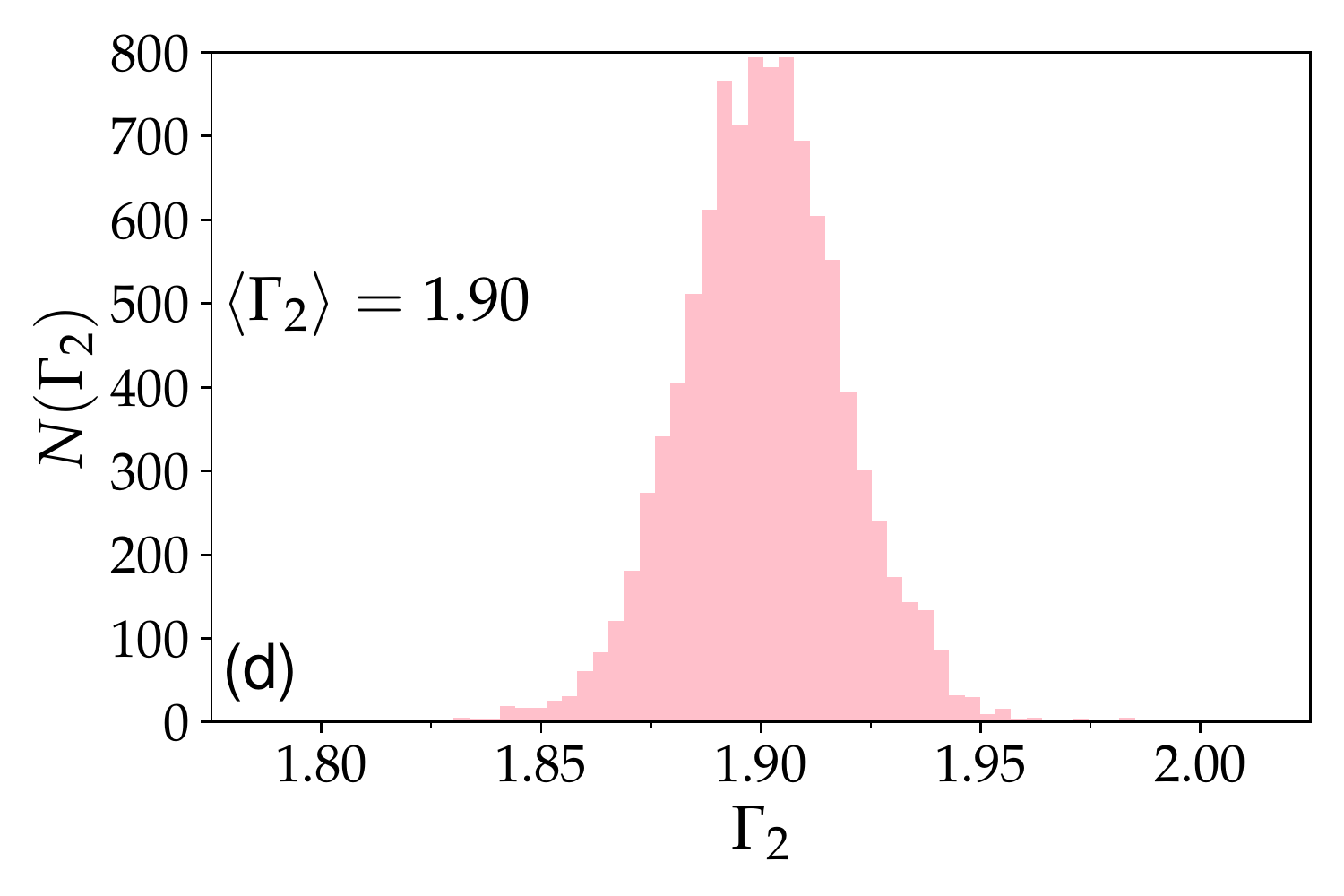}\\
\vspace{-0.7cm}
\includegraphics[scale=0.4]{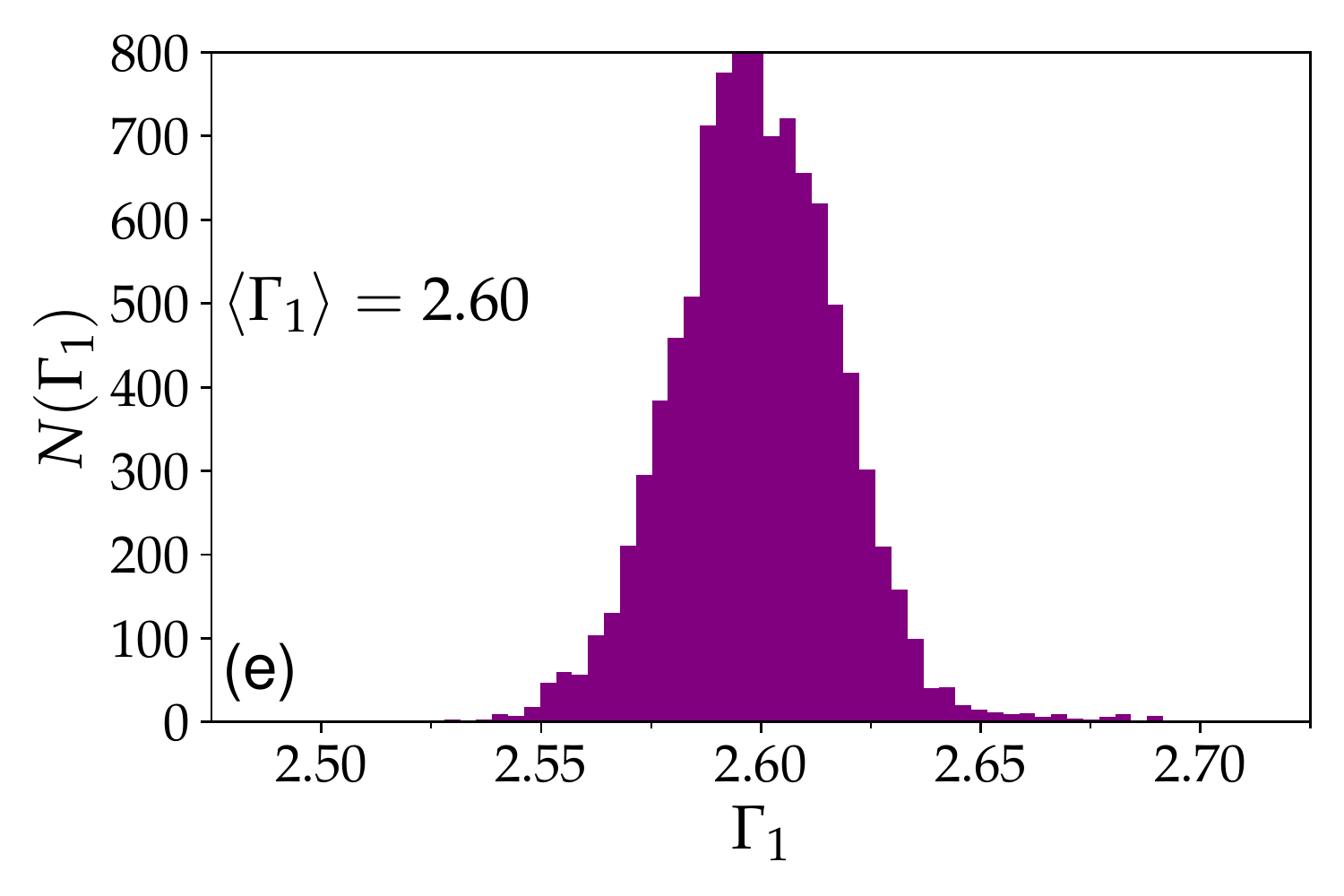}
\includegraphics[scale=0.4]{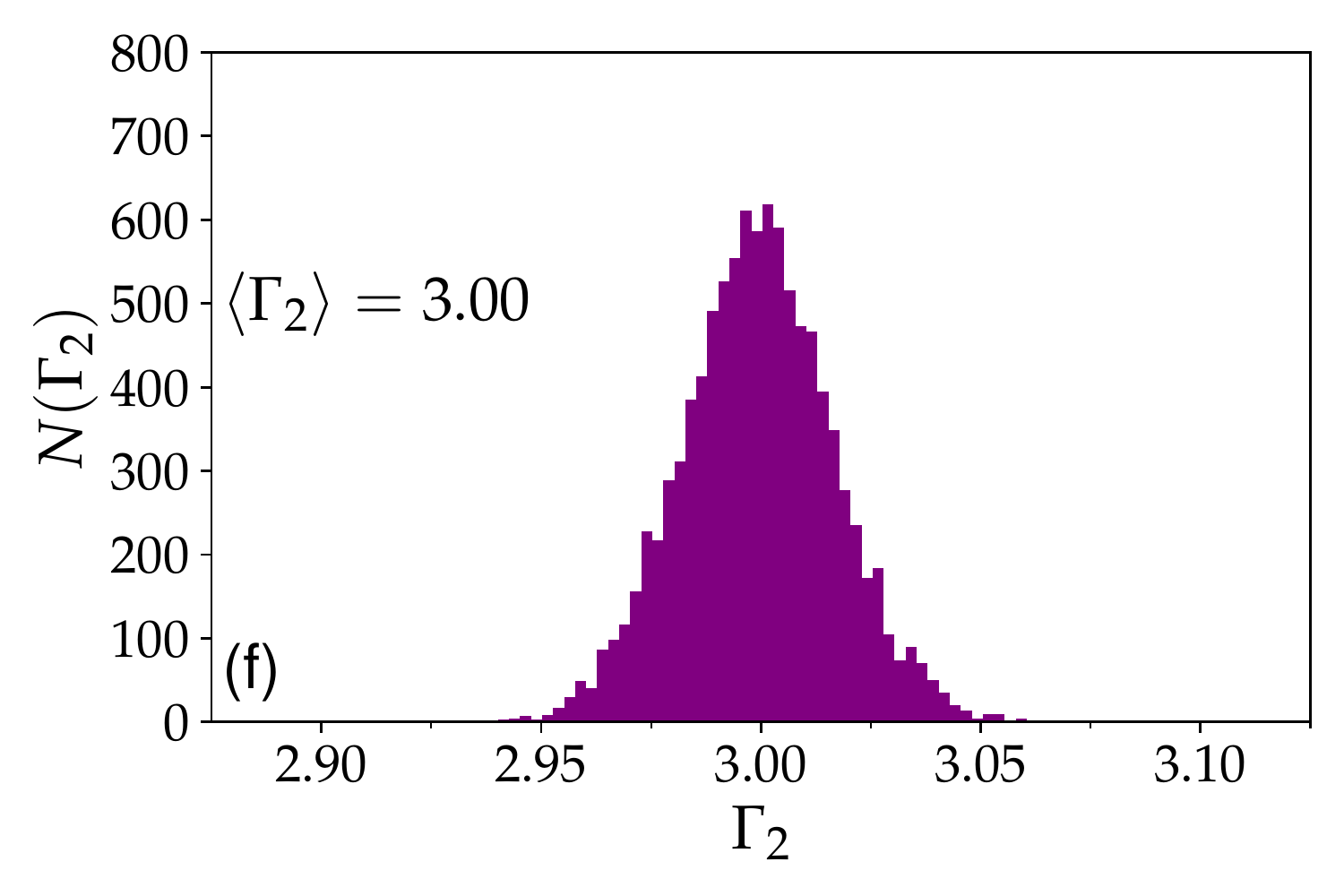}\\
\vspace{-0.7cm}
\includegraphics[scale=0.4]{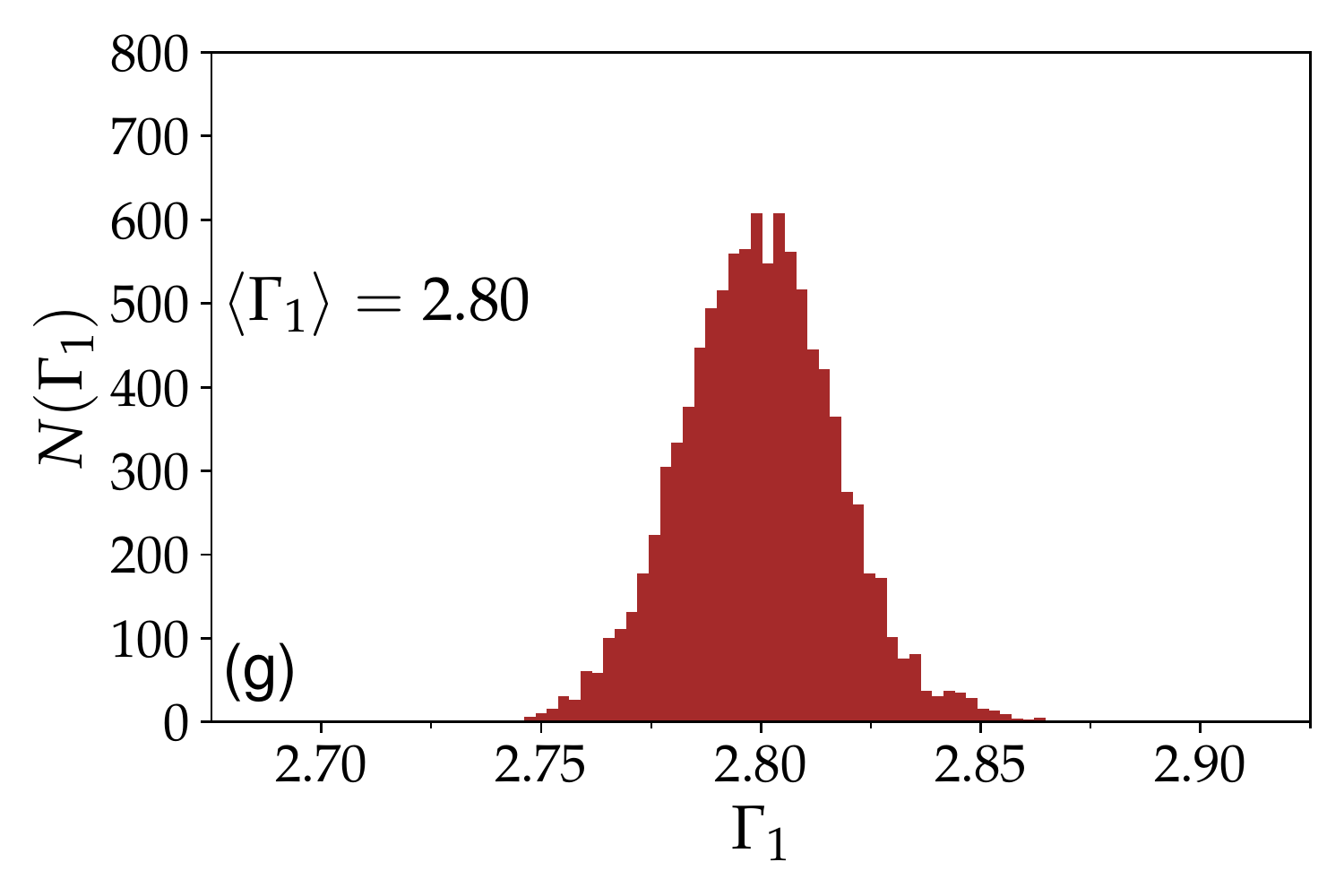}
\includegraphics[scale=0.4]{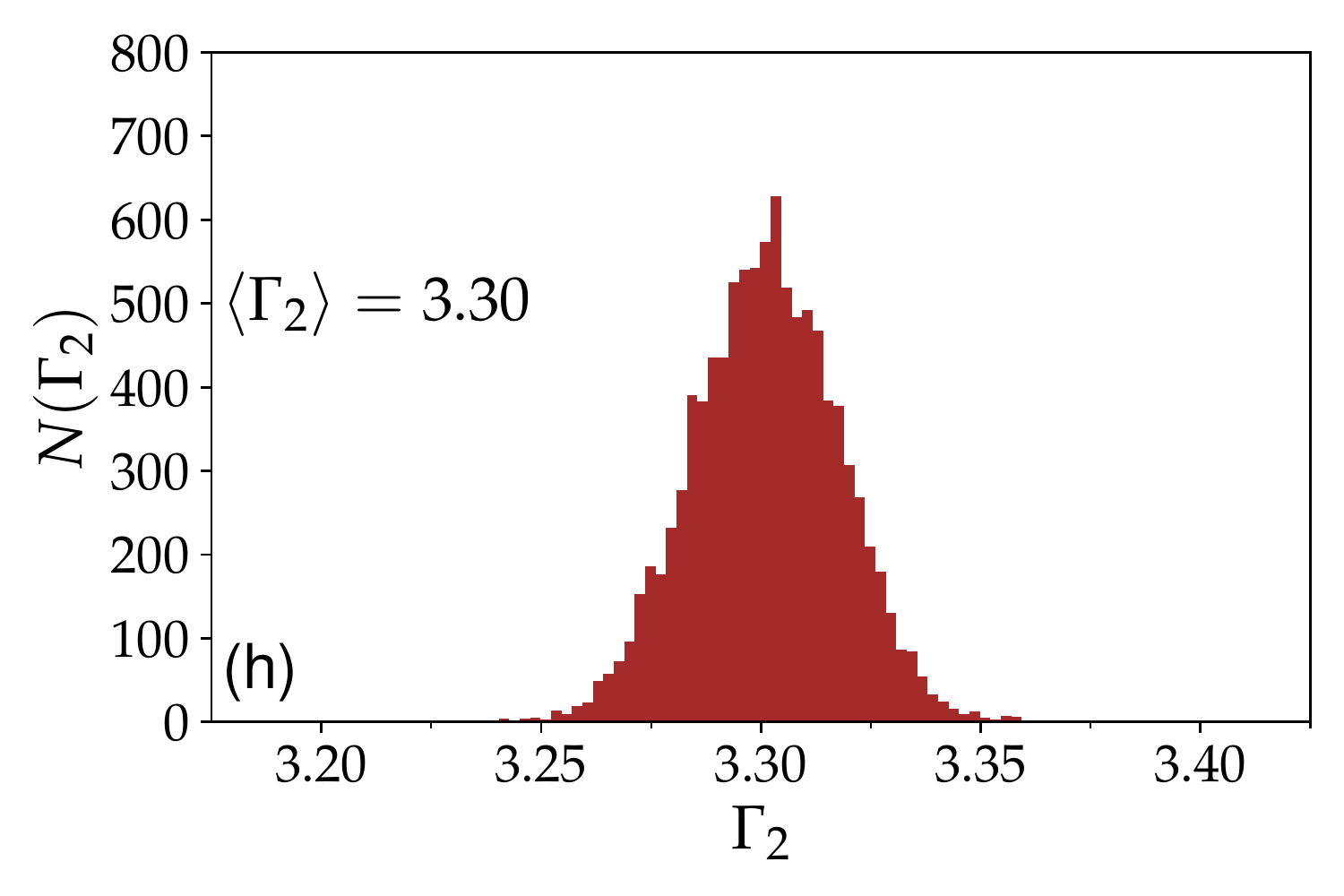}\\
\vspace{-0.7cm}
\includegraphics[scale=0.4]{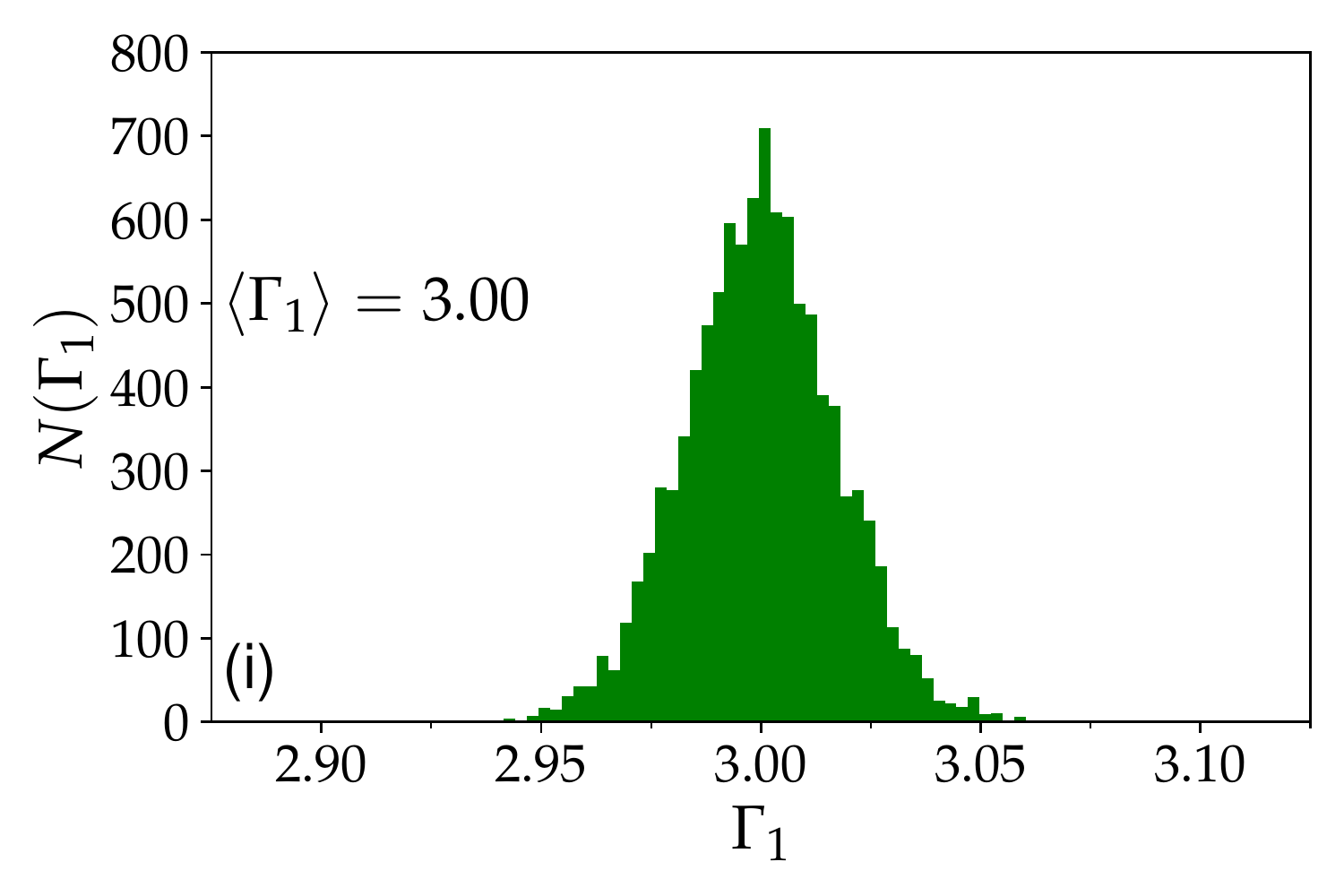}
\includegraphics[scale=0.4]{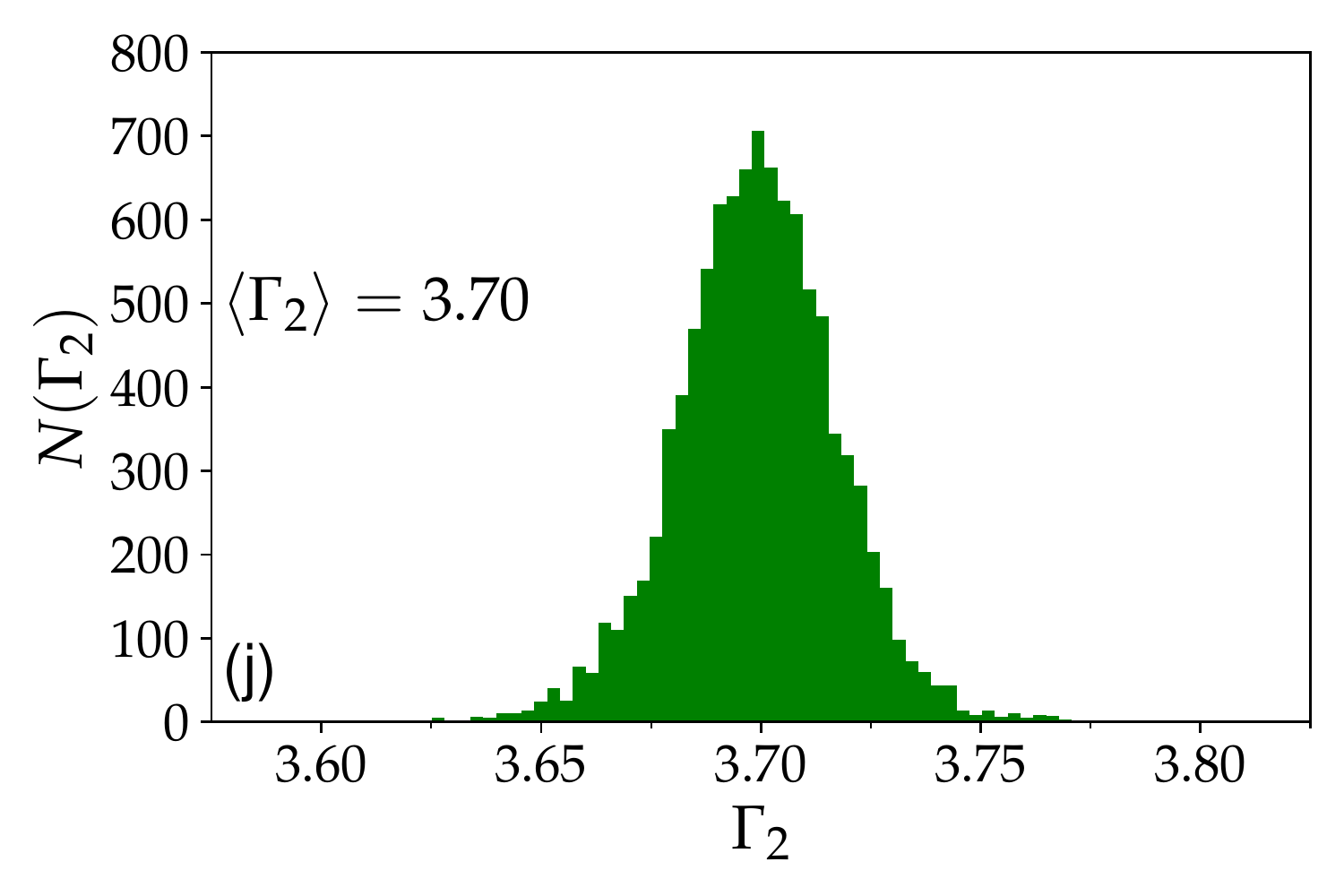}
\caption{\label{MCMC_histograms} Histograms for MCMC sampling for the five models: MD1, MD2, MD3, MD4 and MD5 (from top to bottom). The distributions are obtained with 10000 iterations. The averaged values for each histogram are shown in the inset and match the parameters given Table~\ref{table_MCMC}.}
\end{center}
\end{figure}

Figure \ref{MCMC_histograms} presents the posterior distributions of
values obtained in the MCMC process for the exponents $\Gamma _{1}$
and $\Gamma _{2}$ in the left and right panels, respectively. We
employ $N = 10000$ iterations of the Monte Carlo sampling for each of
the models defined in Table \ref{table_MCMC}. This number of steps
ensure convergence of the posterior distribution presented as
histograms. Figures \ref{MCMC_histograms}(a-b) represent the model MD1
with averaged values for $\langle \Gamma _{1}\rangle = 2.5$ and $
\langle \Gamma_{2}\rangle = 1.8$. The histograms for the parameters
of MD2 is shown in Figs~\ref{MCMC_histograms}(c-d) with $\langle
\Gamma _{1}\rangle = 2.6$ and $ \langle \Gamma_{2}\rangle =
1.9$. The corresponding histogram with averaged values $\langle
\Gamma _{1}\rangle = 2.6$ and $ \langle \Gamma_{2}\rangle = 3.0$
for model MD3 is shown in Figs \ref{MCMC_histograms}(e-f).

The two last subfigures of Fig. \ref{MCMC_histograms}, Figs
\ref{MCMC_histograms}(g-h) and Figs \ref{MCMC_histograms}(i-j), are
associated to models MD4 and MD5 respectively. MD4 has the averaged
values of $\langle \Gamma_1 \rangle = 2.80$ and $\langle \Gamma_2
\rangle = 3.30$. For the MD5, we have the following ones $\langle \Gamma_1 \rangle = 3.00$ and $\langle \Gamma_2
\rangle = 3.70$. The averaged value of each distribution confirms the convergence of the statistical sampling approach.

The MCMC provides the posterior distribution for what we can sample $\Gamma _{1,2}$ values to generate equation of states for each of the model MD$\#$. The EoS is then used to solve the TOV equation to obtain the
mass and radius diagrams. As we are going to see the MD4 and MD5 are
the representations that seem to better reproduce the EoS, according to the observations.

\begin{figure}[h!]
\centering
\includegraphics[width=12.5 cm]{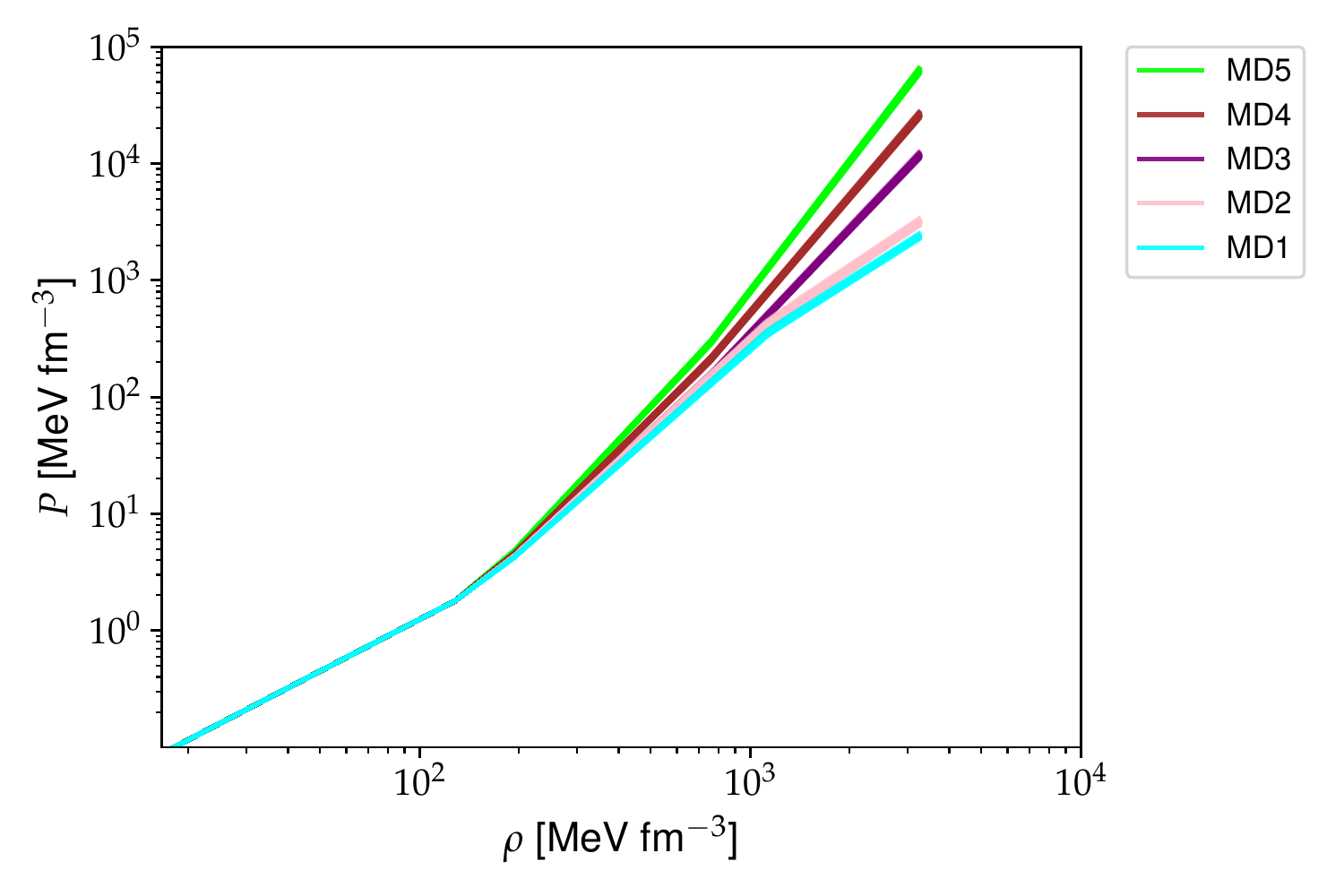}
\caption{The collection of EoS generated for the 5 models defined in Table \ref{table_MCMC}. The description of the regime of low density associated to the nuclear matter saturation point $\rho _{0}$ is kept unchanged for every model. The transition between different polytropic equation can be seen as elbows with large $\Gamma $ values providing highly stiff equation of state.}
\label{eosmcmc}
\end{figure}

In figure \ref{eosmcmc} we have the parametrized EoS for each
model of 5 models studied here. As one can see, at low densities they all converge to a single picture determined by the SLy4 parametrization. At higher densities they diverge from each other, where the MD1 being the
softest model and the MD5 the stiffest one. The parametrizations of
MD1 and MD2 in the third polytrope, $\Gamma_2$, reduces the speed of
sound, which reflects a decrement in comparison with the other models
for higher densities. This feature will be reflected in the mass-radius diagram, as we are
going to see.

\begin{figure}[h]
\centering
\includegraphics[width=14 cm]{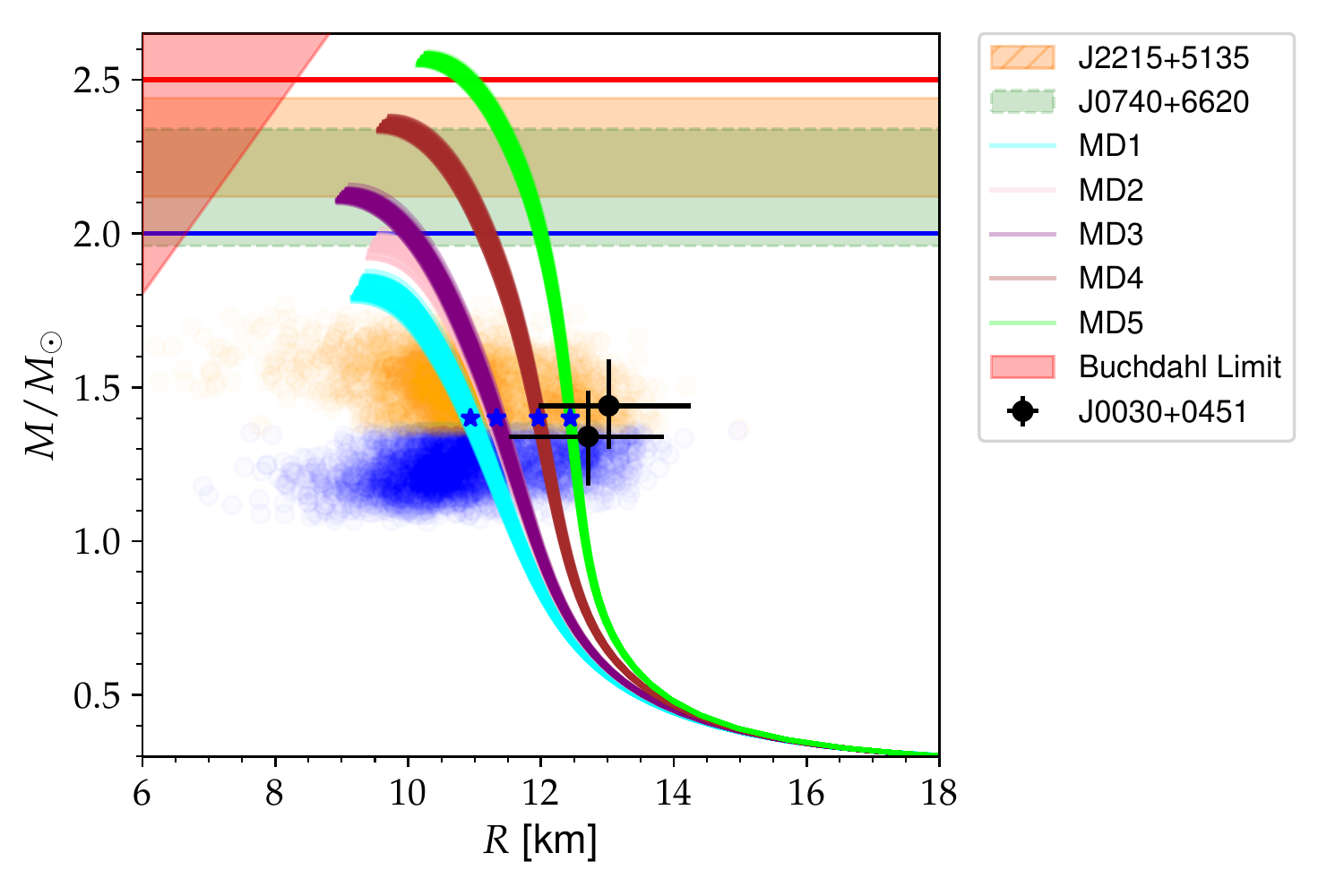}
\caption{Mass radius relationship for the set of all our parametrizations. We have five lines corresponding to the MD$\#$s in table \ref{table_MCMC}. The blue continuous line at $2.0\ M_{\odot}$ corresponds to the two massive pulsars J0348+0432 and J1614-2230. The filled green region represents the pulsar J0740+6620 and the filled dashed salmon region is the pulsar J2215+5135. The red line is the low mass compact object in the binary system GW190414. The black dots with errors bars are the {\it NICER} estimations of PSR J0030+0451. The blue solid stars mark $1.4\ M_{\odot}$ stars.}
\label{MR_all_eos_mcmc}
\end{figure}

In Figure \ref{MR_all_eos_mcmc} we present the mass-radius relation for the
5 sets of our EoS parametrization. They are respectively the models
going from 1 to 5 in table \ref{table_MCMC}. In this figure we also
have the LIGO-VIRGO mass-radius region, in blue and
orange, constrained
by the gravitational wave event GW170817 \cite{ligo/2018,
  ligo/2019}, This constraint was the first one to have a radius
associated with the mass tightly constrained, since the previous
observations for neutron stars were using electromagnetic bandwidth,
which is very difficult to estimate the radius of NS. After this gravitational wave detection, the radius of a
pure nuclear hadronic matter with mass of $1.4\ M_{\odot}$ was
estimated to be $\overline{R}_{1.4}$ = 12.39 km
\cite{most/2018}. Afterwards, this constraint was shifted due
to measurements of {\it NICER} for the pulsar PSR J0030+0451 to two
values: $M \approx 1.44\ M_{\odot}$ and equatorial radius of $R_{\rm
  eq} = 13.02$ km \cite{miller/2019a} and $M \approx 1.24\ M_{\odot}$
and $R_{\rm req} = 12.71$ km \cite{riley/2019}. This information is
highlighted by the black dots with error bars in the mass-radius
diagram. In Fig. \ref{MR_all_eos_mcmc} we present as upper limit of
mass, a lower mass compact object with $2.50-2.67\ M_{\odot}$ in a binary system detected by LIGO-VIRGO \cite{abbott/2020},
this unknown object if it is a NS, will be a breakthrough, since no
nuclear theory for ordinary matter, can explain such a EoS to generate
such a mass in general relativity. Finally, we considered observations of massive pulsars: The extremely massive millisecond pulsar
PSR J0740+6620 with mass of $2.14^{+0.20}_{-0.18}\ M_{\odot}$
\cite{cromartie/2020}; the PSR J2215+5135 with mass $\approx 2.27\
M{\odot}$ \cite{linares/2018}. The measurement of the mass of the source
is not so precise, and if this number is confirmed, the star would be
the most massive neutron star ever detected; the two well-known NS sources J0348+0432 and J1614-2230 \cite{demorest/2010,
  antoniadis/2013} with $M = 2.0\ M_{\odot}$.

As we can see in the mass-radius diagram, the models MD4 and MD5,
brown and lime curves respectively, seem
to better explain the observation. One can see that the curves cross
the region of the {\it NICER} observation. For the mass of $1.4\
M_{\odot}$, the respective radius for
MD4 and MD5 are: $\overline{R}_{\rm MD4} = 11.96$ km and $\overline{R}_{\rm MD5} = 12.44$ km,
the other models for the same mass are more compact and are off the
{\it NICER} data, although they are
still inside the LIGO-VIRGO region. These two approaches also cross the line for $M =
2.0\ M_{\odot}$ pulsars as well as they are inside the region for PSR
J0740+6620 and PSR J2215+5135 mass. Regarding the model MD5, even the
compact object with $2.5\ M_{\odot}$ could be explained with this
parametrization. They have as maximum mass $\overline{M}_{\rm max\ MD4} = 2.35\ M_{\odot}$ and $\overline{M}_{\rm max\
  MD5} = 2.57\ M_{\odot}$ and respective radius $\overline{R}_{\rm MD4}
= 9.63$ km and $\overline{R}_{\rm MD5} = 10.21$ km.

Concerning the model MD3, the purple one, it could explain the massive
pulsars also; however, it starts to get out of the {\it NICER}
constraints. In this region of mass-radius this parametrization mixes
with the model MD2, the pink one. They have similar $\Gamma_1$, which
reflects in the central density when solving the TOV equation and
getting the mass of the star in this region. For the densest region, the model MD2 can barely
reach the two solar mass region. It has a maximum mass of $\overline{M}_{\rm max\
  MD2} = 1.97\ M_{\odot}$ and radius of $\overline{R}_{\rm MD2} = 9.49$ km.
For the model MD1, one cannot have the stars with mass close to two solar
masses, and it lies away of the {\it NICER} measurement.

\section{Bayesian Power Regression Model with Heteroscedastic errors}

In this section, we briefly investigate the potential use of a Bayesian Power Regression model with heteroscedastic errors (BPR-HE) to capture the relationship between the density and pressure. The idea here is to train a model that incorporates the associated variance of a large variety of physics parametrizations. This approach could then be constrained by observational data automatically in a physics informed machine learning strategy. As a preliminary step towards that, we focus on the BPR-HE approach.

\begin{figure}[h!]
\centering
\includegraphics[scale = 0.35]{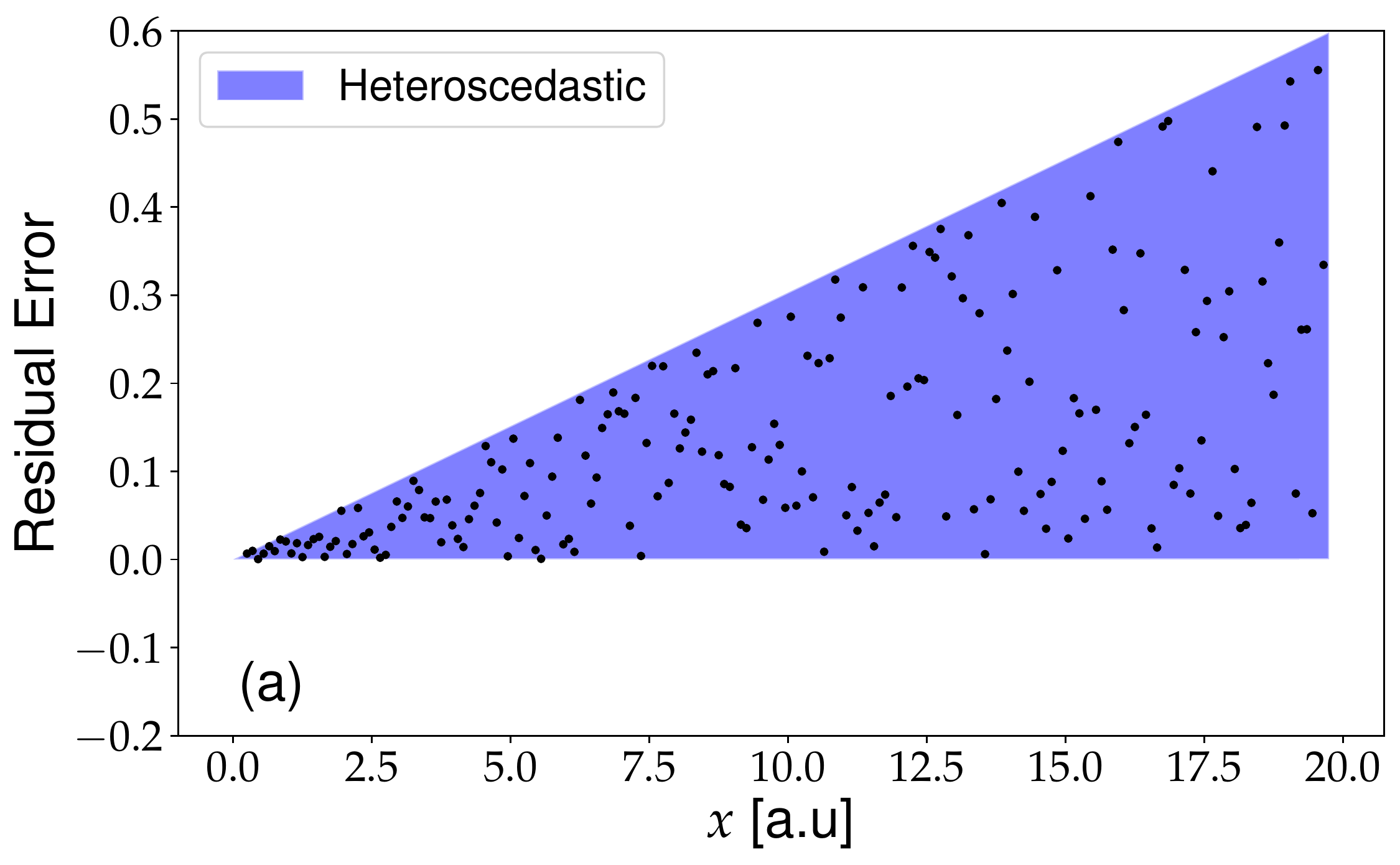}
\includegraphics[scale = 0.35]{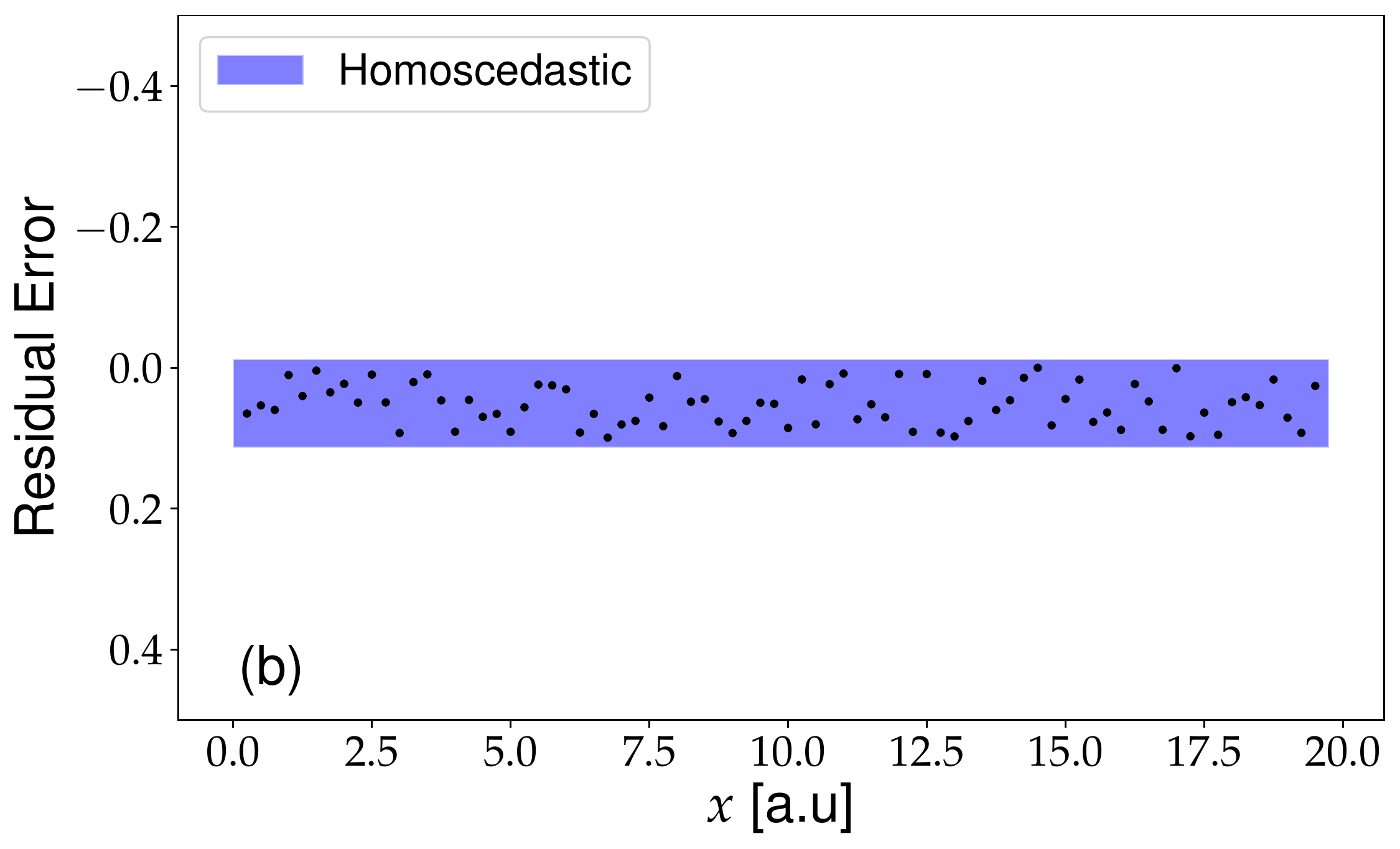}
\caption{Representative example of heteroscedastic (a) and homoscedastic (b) residuals. Notice how the variance of the residuals changes with the value of x for the first, while it remains constant for the second.\label{Heteroscedastic}}
\end{figure}

Power regression is a non-linear regression model that take the form $y = ax^b$, where $y$ is the response variable, $x$ is the prediction variable, and $a$ and $b$ are the coefficients that describe the relationship between $x$ and $y$. The model can be made linear by simply applying a log transformation: $\log(y) = \log(a) + b\log(x)$. Therefore, one can infer the parameters of a non-linear power regression model via a linear model. With that, our corresponding BPR-HE model is defined as
\begin{equation} \label{equation_bpr_he}
    \begin{split}
    \log(p^i) & \sim \text{Normal}(\alpha\cdot \log(\rho^{i}) + \beta, s_m \cdot \log(\rho^{i}) + s_b), \hspace{1cm} \forall i=1,..., N. \\
    \alpha & \sim \text{Normal}(\gamma_1, \gamma_2) \\
    \beta & \sim \text{Normal}(0, \gamma_3) \\
    s_m & \sim \text{HalfCauchy}(\gamma_4) \\
    s_b & \sim \text{HalfCauchy}(\gamma_5)
    \end{split}
\end{equation}

\noindent where $\gamma_*$ are a set of hyperparameters that are
specified by the user. In our experiment, we set all $\gamma$ to 1. N = 65
is the total number of equation of states taken from the LIGO {\it Lalsuite} \cite{lalsuite} library
 and used as data set. Essential to
the model is the dependence of the standard deviation of the residual
to the density variable $\rho$. This is necessary as the ensemble of EoS from LiGO have an increasing pressure variance with respect to densities. Residuals with varying variance is known as heteroscedastic. Figure~\ref{Heteroscedastic} shows an illustrative example of heteroscedastic errors and homoscedastic errors, which is assumed in classical linear regression models.

We used the Numpyro probabilistic programming language \cite{bingham2019pyro} to implement the model~\ref{equation_bpr_he}. We inferred the values of the unknown parameters in our model ($\alpha$, $\beta$, $s_m$, and $s_b$) by running MCMC using the No-U-Turn Sampler (NUTS) \cite{hoffman2014no} for 10000 warm-up samples and then collected 1000 posterior samples to represent our model parameters posterior distribution.

\begin{figure}[h!]
\centering
\includegraphics[width=12.5 cm]{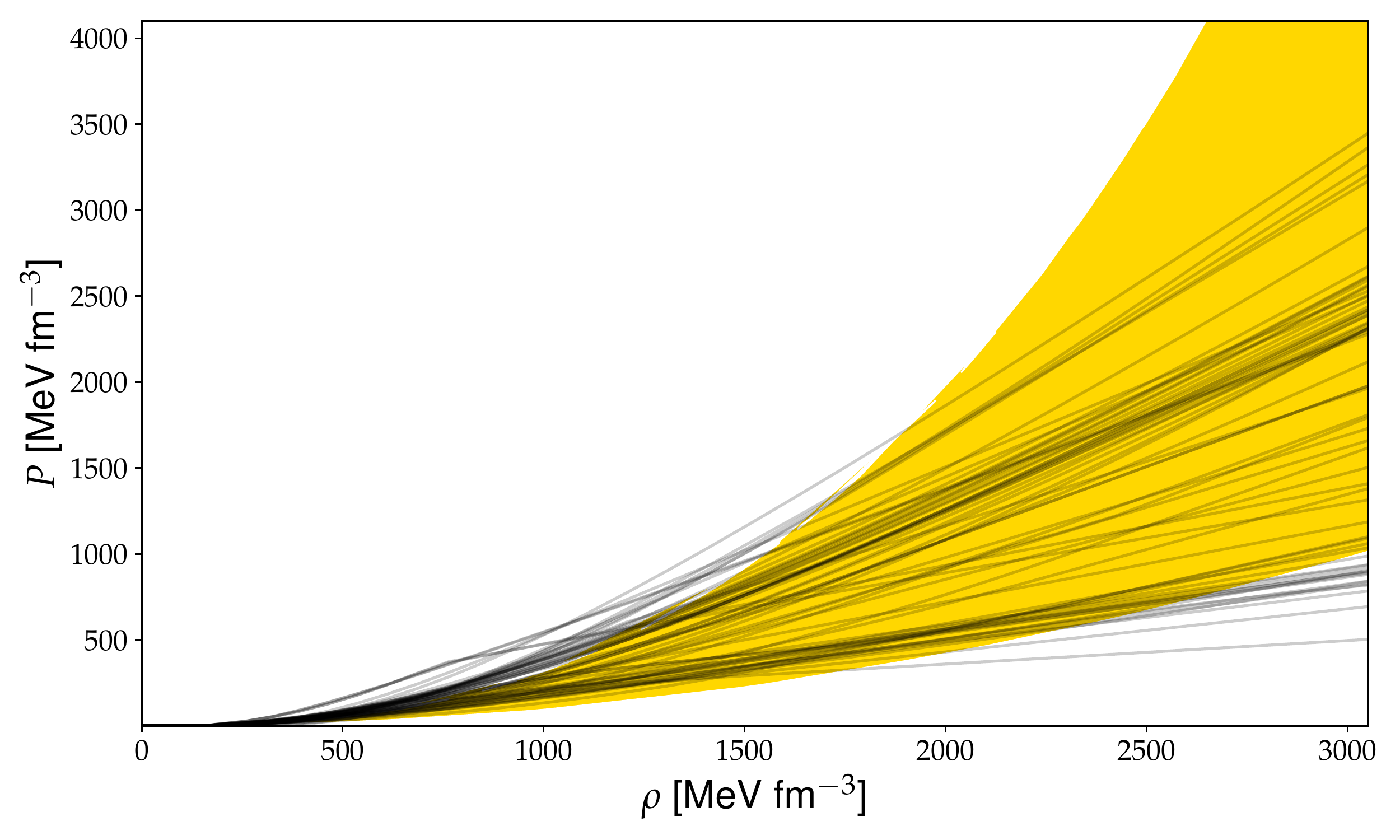}
\caption{Bayesian Power Regression model heteroscedastic errors. Black solid lines are the 65 EoS from the LIGO {\it Lalsuite} \cite{lalsuite} data set, while the yellow ones are posterior samples generated by the BPR-HE model.}
\label{posterior_samples}
\end{figure}

Figure~\ref{posterior_samples} shows posterior samples of the BPR-HE model (in yellow), and the EoS from LIGO. Notice that the model captures the increasing variance of the pressure as the density increases. This is due to the heteroscedastic errors in the model. The BPR-HE provides smooth functions as opposed to the piecewise polytropic approach described in the previous section. The abrupt transition points of the MD$\#$ models can reduce the accuracy of the description of local speed of sound, a problem already discussed in \cite{read/2009}.

From Figure~\ref{posterior_samples}, we observe that the BPR-HE model is a promising approach to be used to model the relation of density and pressure. This information can be used to estimate uncertainties of the maximum radius and mass of the neutron stars via TOV equations. We left the use of BPR-HE generated EoS into the TOV equation for future works and have restricted this work to a feasibility analysis of such approach.

\section{Conclusion and perspectives}
In this paper, we have studied parameters of piecewise polytropic equations that are representative of an equation of state modeling the interior structure of neutron stars. Polytropic piecewise models with few parameters are known to be good approximations to modern theoretical EoS and are able to reproduce global features of neutron stars such as mass, radius, moment of inertia and so on (e.g. see Ref. \cite{read/2009, Kurkela-2014, raithel/2016, steiner/2016, raaijmakers/2018, Miao_2021}). Commonly used in literature, this phenomenological approach is applied in a broad research context from numerical solution of rotating relativistic stars/merger simulations \cite{endrizzi/2016, maione/2016, east/2019} to modified gravity \cite{astashenok/2013, anderson/2019a, lobato/2020, odintsov/2021, odintsov/2021a} studies. Also, recently, they have successfully been applied to constrain the dense matter equation of state of neutron stars supported by observations \cite{raaijmakers/2019, miller/2019b, raaijmakers/2021, Miao_2021}.

In this work, we have connected three politropes resulting in 5
different schemes globally adjusted. We performed an analysis that
have accounted for different astronomical observational sources, in a
joint constraint. We have performed a Bayesian analysis of piecewise
equations and then Markov Chain Monte Carlo (MCMC) strategies were we
employed to access the variability of our models to constrain the
representation of the EoS compared with observations of neutron
stars. Furthermore, we have obtained 2500 mass-radius diagrams within
the 5 different model schemes. The massive stars $M > 2 M_{\odot}$ and
stars with mass around $M \approx 2 M_{\odot}$ can be explained with
combinations of adiabatic indices of $2.8 \lessapprox \Gamma_{1,2}
\lessapprox 3.7$, as in the case of MD4 and MD5. Increasing this
exponent even further will stiff more the EoS and result in
unrealistic stars.

From our adopted schemes, the two models, MD4 and MD5, can represent very well stars for mass around $1.4\ M_{\odot}$ and radius of $\approx 12$ km, i.e., the mass-radius observational region of LIGO-VIRGO binary NS merger and the PSR J0030+0451 constrained by the {\it NICER} experiment. The two models can also explain massive pulsars with mass above $2.0\ M_{\odot}$ as the two pulsars J0348+0432 and J1614-2230. One of the models, MD5, can even explain an unknown object with mass of $2.5\ M_{\odot}$ in a binary system, detected by LIG0-VIRGO. This model has a maximum mass of $\overline{M} = 2.57\ M_{\odot}$. As one can notice, the model MD5 yields almost the same radius for different masses, almost a limit for the polytropic exponent.

The parameters found here for the piecewise equations that represent modern EoS, will be used to constraint nuclear models, i.e., the parameters of many-body models. In previous works \cite{lobato/2022, lobato/2022a}, we analyzed correlations in the microphysics of many EoS and in the global properties. We have studied two separated spaces, and now we are able to bring these two complementary studies together in a full picture and by means of statistical and machine learning tools, shading light in the path to understand the EoS of neutron stars.

Before closing, we would like to comment on some challenges and remarks on modeling EoS coming from many nuclear models with different parametrizations. Statistical models, such as regression model with heteroscedastic errors, for example, has the potential to best represent a set of different physics included in a variety of equation of states. The Bayesian Power Regression model with heteroscedastic errors (BPR-HE) is a flexible model, but we faced difficulties with it due to the nature of the data. In the model, the variance of the errors varies linearly with the density value, which might not be appropriate, as the $s_m$ parameter has shown to be very sensitive and difficult to infer. We had to resort to a forceful (informative) prior to stabilize the inference. Another point is that using a single power regression model to describe all EoS might be too restrictive, given the diversity of physical models. We believe that a mixture regression model, composed of several power regressors, will bring more flexibility. These points will be the focus of future research steps, as well the tension brought by the Lead Radius EXperiment (PREX-2) results with astronomical data.


\vspace{6pt}




\section*{Funding}
This research was partly funded by U.S. Department of Energy (DOE)
under grant DE--FG02--08ER41533 and to the LANL Collaborative Research
Program by Texas A\&M System National Laboratory Office and Los Alamos
National Laboratory. Also, partly funded by UNIANDES University. The
work at Brookhaven National Laboratory was sponsored by the Office of
Nuclear Physics, Office of Science of the U.S. Department of Energy
under Contract No. DE-AC02-98CH10886 with Brookhaven Science
Associates, LLC.

\bibliographystyle{unsrt}
\bibliography{lib.bib}


%


\end{document}